\documentclass[aps,
		prd,
		reprint,
		twocolumn,
		superscriptaddress,
		shortbibliography,
		nofootinbib,
		floatfix,
		notitlepage
		]{revtex4-1}
\pdfoutput=1

\usepackage{slashed}
\usepackage{mathrsfs}
\usepackage{amsmath}
\usepackage{amssymb}
\usepackage{revsymb}
\usepackage{graphicx,accents}
\usepackage{graphicx,epstopdf}
\usepackage{mathrsfs}
\usepackage{bm}
\usepackage{psfrag}
\usepackage{hyperref}
\hypersetup{colorlinks=true,citecolor=Red,urlcolor=Red}
\usepackage{bbm}
\usepackage{bbold}
\usepackage{tabularx}
\usepackage{graphicx}
\usepackage[normalem]{ulem}
\usepackage[usenames,dvipsnames]{xcolor}

\usepackage[usenames,dvipsnames]{xcolor}
\usepackage{multirow,tabularx}
\usepackage{xcolor,pifont}
\newcommand*\colourcheck[1]{%
  \expandafter\newcommand\csname #1check\endcsname{\textcolor{#1}{\ding{52}}}%
}

\newcommand{\be}{\begin{equation}}
\newcommand{\ee}{\end{equation}}
\newcommand{\bea}{\begin{eqnarray}}
\newcommand{\eea}{\end{eqnarray}}

\newcommand{\trm}[1]{\textrm{#1}}

\newcommand{\vphi}{\varphi}

\newcommand{\LCperp}{{\scriptscriptstyle \perp}}

\newcommand{\ud}{\mathrm{d}}

\newcommand{\vtheta}{\vartheta}

\definecolor{bk1}{RGB}{0,200,100}

\newcommand{\ST}[1]{\textcolor{black}{#1}}

\begin{document}
\title{Polarization-dependent Interference in Nonlinear Compton Scattering}
\author{Zu-dong Zhao}
\affiliation{College of Physics and Optoelectronic Engineering, Ocean University of China, Qingdao, Shandong, 266100, China}

\author{Suo Tang}
\email{tangsuo@ouc.edu.cn}
\affiliation{College of Physics and Optoelectronic Engineering, Ocean University of China, Qingdao, Shandong, 266100, China}
\affiliation{Engineering Research Center of Advanced Marine Physical Instruments and Equipment, Ministry of Education, Qingdao, Shandong, 266100, China}
\affiliation{Qingdao Key Laboratory of Optics and Optoelectronics, Qingdao, Shandong, 266100, China}


\begin{abstract}
We investigate the phase interference effects in the nonlinear Compton scattering via the collision between an high-energy electron and the laser in the intermediate intensity region,
and reveal that the importance of interference effects on the scattered photons depends sensitively on the relative polarization of the scattered photons and laser pulse.
For scattered photons polarized parallel to the laser field, the effective distance between two interfered phase points is much larger than that for perpendicularly polarized photons.
To signalize the phase interference effect in potential experiments, we introduce the double-pulse scenario for the scattering process, and show that the interference between two phase-separated pulse could significantly modulate the energy-momentum distribution of the scattered photons,
leading to the reshaping of the scattered photons' energy spectrum.
By narrowing the angular acceptance of the scattered photons, the spectral modulation can be amplified and sustained for an experiment-realizable phase separation between two pulses.


\end{abstract}
\maketitle
%
%
\section{Introduction}
Compton scattering, as one of the fundamental processes in light-matter interactions, has served as a cornerstone in quantum electrodynamics (QED) since 1923~\cite{PhysRev.21.483}.
It not only validates the particle nature of light~\cite{Evans1958}, but also lays the groundwork for the generation of hard-photon source~\cite{HAJIMA201635}.
Using a weak laser pulse with normalized intensity $\xi\ll1$, the scattering process exhibits linearity and perturbation as elastic collision between a single laser photon and a high-energy electron with the cross-section given as Klein-Nishina formula~\cite{landau4,greiner2008quantum}.
The current breakthrough of laser technique has brought the intensity to the highly relativistic regime with \mbox{$\xi\gg1$}~\cite{lee2018exploration,SULF}, in which the acceleration of electrons within one laser cycle can improve their energy exceeding the static energy of electron~\cite{olofsson2023prospects}.
Colliding with such strong lasers, the linear and perturbative QED description of the Compton scattering fails as electrons are strongly dressed by the laser fields, and plenty of laser photons can contribute simultaneously to the scattering process, \emph{i.e.}, nonlinear Compton scattering (NLC)~\cite{RMP2012Piazza,RMP2022_045001,Fedotov:2022ely}.

The NLC process has been theoretically investigated via the QED scattering theory~\cite{nikishov64,seipt2017volkov}, semiclassical operator method~\cite{PhysRevD.90.125008,PRD013010}, and particle-tracking Monte Carlo simulations~\cite{gonoskov15,PoP093903,TANG2024139136} in various field configurations: monochromatic field~\cite{kibble64,serbo04,harvey09,Seipt_2013}, constant crossed field~\cite{king19a,PiazzaPRA2019,SeiptPRA052805} and laser field with finite duration~\cite{boca09,PRL063903,heinzl10b,PRA032106,seipt11,MRE0196125}, bi-frequency~\cite{PhysRevA.100.061402} and double-pulse structure~\cite{PhysRevD.90.125008,ILDERTON2020135410}.  
Experimentally, the NLC was first observed in the perturbative regime with up to four laser photons converted to a single $\gamma$ photon by a high-energy electron of $46.6~\trm{GeV}$ colliding with a laser pulse of $\xi=0.6$~\cite{bula96,bamber99},
and recently was tested nonperturbatively as hundreds of laser photons were scattered into a single $\gamma$ photon in the collision between laser-wakefield accelerated multi-GeV electron beams and laser pulses of $\xi>10$~\cite{NPAllopticalNLC}.
The later experimental scenario has also been used to verify the quantum nature of radiation recoil induced by the scattering process~\cite{PRX011020,PRX031004}.

The upcoming laser-particle experiments, such as LUXE at DESY~\cite{Abramowicz:2021zja} and E320 at SLAC~\cite{slacref1,E320_2021}, are designed to measure QED processes in the transition regime from the perturbative to nonperturbative with high resolution~\cite{Macleod2022},
using electron beams of $E_{p}\sim O(10)~\trm{GeV}$ and laser pulses in the intermediate intensity regime of $\xi\sim O(1)$.
With these parameters, the formation length of the scattering process could be comparable to the scale of the field variation~\cite{ritus85}, and therefore, the quantum interference between different phase points in the field would play a critical role in the scattering process.
One of the typical characters resulting from these phase interference, is the harmonic structure in the spectrum of the scattered photons when the laser pulse has long-enough duration~\cite{Seipt_2013,BenPRA2020}.
This harmonic structure can, on one hand, be associated with the net number of laser photons joined in the scattering process, and on the other hand, be attributed to the constructive/destructive interference between different phase points in the scale of laser wavelength~\cite{LMA063110}.
Previous studies revealed that clear harmonic structure can only appear in the spectrum of the scattered photons in the polarization state parallel to the laser polarization~\cite{BenPRA2020}, but not in the photon's spectrum in the perpendicular polarization.
This implies the dependence of the phase interference on the relative polarization between the laser and scattered photons.

In this paper, we intend to figure out the polarization-dependence of this phase interference in the NLC process,
and find out measurable interference-induced signals with realistic experimental setups.
To this objective, we resort to the collision scenario between a high-energy electron and a linearly polarized laser with double pulses.
The similar scenario has been employed in Ref.~\cite{ILDERTON2020135410},
in which single-cycle pulses were used for observable interference rings in the momentum distribution of the scattered photons,
and claimed that for relatively long laser pulses, the interference signals would be smoothed out for the detector with reasonable resolution.
Here, we show that the interference between two multi-cycle pulses can not only modulate the energy-momentum distribution of the scattered photons, but also reshape the harmonic spectrum of the scattered photons.

The paper is organised as follows.
In Sec.~\ref{Sec_2}, we introduce briefly the theoretical model used for the investigation of NLC and reveal the relation between the phase interference and the harmonic structure in the photon's spectrum. 
In Sec.~\ref{Sec_3}, we analyze the interference between two phase-separated pulses and give potential interference-induced signals in experiments, and at the end conclude in Sec.~\ref{Sec_4}.
Throughout our discussion, the natural units $\hbar=c=1$ is used.


\section{THEORETICAL MODEL}~\label{Sec_2}
We proceed our discussion with the parameters in the upcoming laser-particle experiments, such as LUXE~\cite{Abramowicz:2021zja}, simplify the laser field as a plane wave with the scaled vector potential $a^{\mu}(\phi) = |e|A^{\mu}(\phi)$ propagating with the wavevector $k^{\mu}=\omega_{0}(1,0,0,-1)$ and carrier frequency $\omega_{0}=1.55~\trm{eV}$,
and calculate the energy- and momentum-distributions of the photons scattered by a high-energy electron with the energy parameter $\eta=k\cdot p/m^{2}$,
where $\phi=k \cdot x$ is the phase of laser pulse, $|e|$, $m$ and $p^{\mu}$ are respectively the absolute charge, mass and momentum of the electron.

We employ direct QED calculations for the NLC process in order to resolve the phase interference in the scale close to and/or much longer than the laser wavelength. 
These interference are much beyond the local approximations: the locally constant field approximation (LCFA) is blind to the phase interference~\cite{PRA013822} and the locally monochromatic approximation (LMA) can only resolve the interference in the scale of the laser wavelength~\cite{LMA063110}.
The triple-differential probability for the scattered photon linearly polarized in the plane perpendicular to the laser propagating direction 
can be formulated as~\cite{tang2020highly}
\begin{align}
\frac{\ud^{3}P_{x,y}}{\ud s \ud^2\bm{r}}&=\frac{\alpha}{(2\pi\eta)^2} \frac{s}{1-s}\left(\frac{s^{2}}{8(1-s)}|w_{0}|^{2} + \left|w_{x,y} \right|^{2}\right) \,,
\label{AllPhoton}
\end{align}
where $\alpha$ is the fine-structure constant,
\mbox{$s=k\cdot \ell/k\cdot p$}, the fraction of the lightfront momentum taken by the scattered photon with momentum $\ell^{\mu}$ from the incoming electron, and {$\bm{r}=r(\cos\psi,\sin\psi) = \bm{\ell}_{\LCperp}/sm-\bm{p}_{\LCperp}/m$}, photon momenta perpendicular to the laser propagating direction, scaling the angular spread \mbox{$\ell_{\LCperp}/sm = m\eta/\omega_{0}\tan(\theta/2)$} along the incoming electron momentum. 
$|w_{0}|^{2} = 2|\bm{F}|^2- S^{\ast}I - SI^{\ast}$ is the unpolarized part relating to the square of the laser field,
and $\bm{w}(\phi)=(w_{x},w_{y}) = \bm{r}I - \bm{F}$ denotes the part in the amplitude level relating to the polarization of laser field $\bm{F}$.
The polarized/unpolarized parts are calculated in term of the integrals~\cite{ILDERTON2020135410}, as
\begin{subequations}
\begin{align}
	I&=\int\!\ud\phi \, \left[1- \ell\cdot \pi_{p}(\phi)/\ell\cdot p \right]~e^{i\Phi(\phi)}\,,\\
	\bm{F} &= \frac{1}{m}\int\!\ud \phi ~ \bm{a}(\phi)~e^{i\Phi(\phi)}\,, \\
	S &= \frac{1}{m^2}\int\!\ud\phi~ \bm{a}^{2}(\phi)~e^{i\Phi(\phi)}\,,
\end{align}
\label{Eq_cal}
\end{subequations}
$\!\!\!\!$over the laser phase $\phi$, where $\Phi(\phi)=\int^{\phi}_{\phi_{i}}\ud\phi'\ell\cdot \pi_{p}(\phi')/[(1-s)\eta m^{2}]$ is the phase integral starting from $\phi_{i}$ at which the laser turns on, and \mbox{$\pi^{\mu}_{p}= p^{\mu}+a^{\mu}- k^{\mu}[p\cdot a/k\cdot p + a^{2}/(2k\cdot p)]$} is the electron's instantaneous momentum in a plane wave. In~(\ref{AllPhoton}), the spin of the recoiled (initial) electron has been summed (averaged).

The contribution of the interference between different field points can be clearly seen as the cross terms between different parts of the phase integrals~(\ref{Eq_cal}) when calculating their modulus squares in the scattering probability~(\ref{AllPhoton}).
The modulus square of the integrals in~(\ref{Eq_cal}) would bring two integrals over the laser phase as $|\int\ud \phi~(\cdots) |^{2} = \int\ud\phi\int\ud \phi' (\cdots)$ into the scattering probabilities.
To manifest the effective distance between two interfered phase points, we write the two phase integrals as
\[\int\ud\phi\int\ud \phi' \to \int\ud \vphi\int \ud\vtheta, \]
where $\vphi=(\phi+\phi')/2$ is the average phase, and $\vtheta=\phi-\phi'$ is the so-called interference phase~\cite{PRL044801,seipt2017volkov}, usually used as the formation length of the scattering process~\cite{ritus85}, and can be used to quantify the phase distance between two interfered field points.

In the LCFA algorithm, widely employed in the ultrahigh intensity regime of $\xi\gg1$~\cite{gonoskov15,RMP2022_045001},
the effects of the phase interference are excluded under the assumption of $\vtheta\ll1$, and the average phase $\vphi$ is regarded classically as the local phase of the field, at where the photon is emitted~\cite{BenPRA2020}.
In our concerned intermediate intensity regime, $\vphi$ and $\vtheta$ are the quantum phase for the scattering process and actually unobservable in experiments.
Below, we employ phase distributions of the scattering probabilities to analyze the polarization-dependence of the phase interference, and resort to the energy-momentum distributions of the outgoing photons as potential experimental observables to signalize the interference effects.

\subsection{Interference and harmonic in energy spectra}

\begin{figure}[t!!!]
 \includegraphics[width=0.49\textwidth]{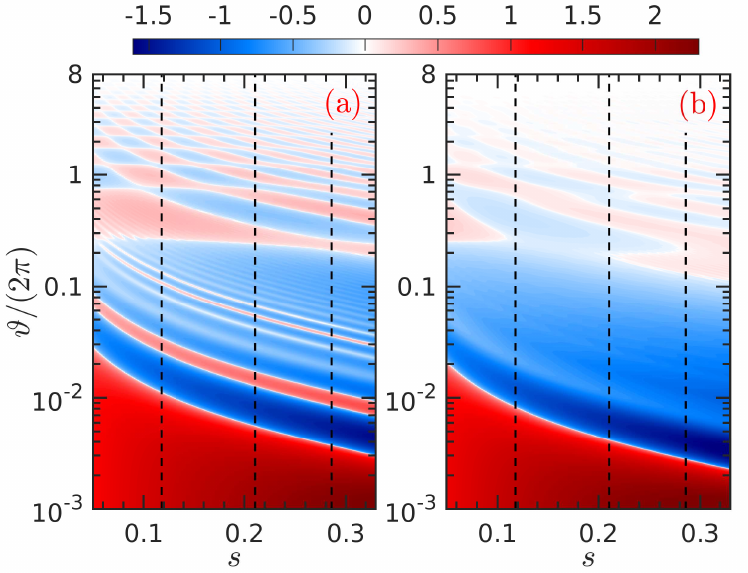}
\caption{Effective length of the interference phase $\vtheta$ with the change of photon energy $s$: (a) $(\ud^{2}P_{x}/\ud s \ud \vtheta)^{1/3}$ and (b) $(\ud^{2}P_{y}/\ud s \ud \vtheta)^{1/3}$ for photons polarized parallel and perpendicularly to the laser field, respectively.
The photons are scattered by a high-energy electron with $\eta=0.1$ colliding with a laser linearly polarized in the $x$ direction with the intensity $\xi=1$ and $N=8$ cycles.
The black dashed lines denote the locations of the harmonic edges at $s_{n,e}=2n\eta/(2n\eta+1+\xi^{2}/2)$ for respectively $n=1,~2~,3$ from left to right.
The details of the pulse functional form is given in the main text.}
\label{Fig_Ds_phi}
\end{figure}

\begin{figure}[t!!!]
 \includegraphics[width=0.48\textwidth]{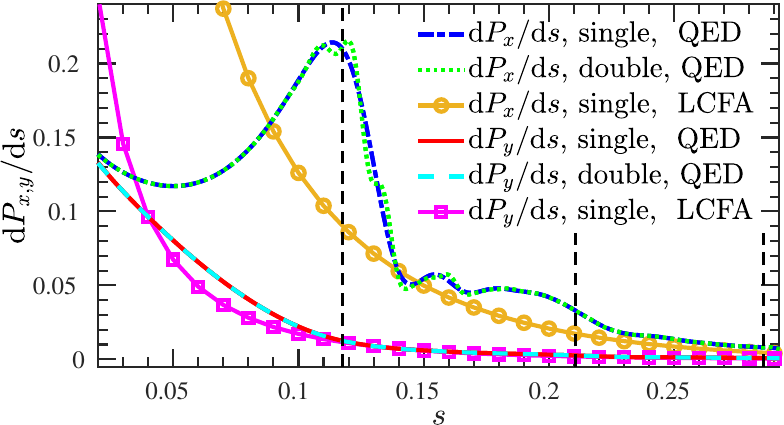}
\caption{Energy spectra of the polarized photons scattered by a high-energy electron colliding with a linearly polarized laser with, respectively, a single pulse and double pulses, calculated with the full QED Eq.~(\ref{AllPhoton}) and LCFA algorithm~\cite{BenPRA2020}.
For the double-pulse case, the two pulses have the exactly same functional form and polarization as the single-pulse laser and are separated with the phase gap $\Delta=3.5\pi$.
For comparison, the energy spectra in the double-pulse case are given by dividing a factor of $2$.
The black dashed lines denote the locations of the first three harmonic edges at $s_{n,e}=2n\eta/(2n\eta+1+\xi^{2}/2)$.
The same other parameters as in Fig.~\ref{Fig_Ds_phi} are used.
}
\label{Fig_spec}
\end{figure}

Fig.~\ref{Fig_Ds_phi} presents the distribution of the interference phase $\vtheta$ for differently polarized photons scattered by a high-energy electron with $\eta=0.1$, corresponding to the energy of $8.4~\trm{GeV}$, in the head-on collision with a \emph{single}-pulse laser as $a^{\mu}(\phi)=m\xi\epsilon^{\mu}_{x}f(\phi)$ with the functional profile $f(\phi)=\cos(\phi)\cos^2[\phi/(2N)]$ in $\left|\phi\right|<N\pi$ and $f(\phi)=0$ otherwise, where $N=8$, $\xi=1$ is the normalized field amplitude, and $\epsilon^{\mu}_{x}=(0,1,0,0)$ denotes the field polarization in the $x$-direction.
As shown, the interference phase could be much larger than one laser cycle, i.e., $\vtheta\gg2\pi$, and further increasing of $\vtheta$ would suppress the emission.
With the increase of photon energy $s$, the effective length of $\vtheta$ becomes much smaller, corresponding to the better performance of the LCFA in describing the emission of higher-energy photons~\cite{PRA012505} (as also shown in Fig.~\ref{Fig_spec}).
The effective region of the interference phase depends also on the photon polarization: the value of $\vtheta$ for photons polarized parallel to the background field in Fig.~\ref{Fig_Ds_phi} (a) can extend to the same level of the laser duration as $\vtheta\sim 2N\pi$, which is much broader than that for the photons polarized perpendicularly to the laser field in Fig.~\ref{Fig_Ds_phi} (b).
This implies that the interference-induced effect for the perpendicularly polarized photons should be much weaker than that for the photons in the polarization parallel to the laser background.
This difference between the differently polarized photons can be simply seen in (\ref{AllPhoton}) as $\int \ud\psi (\left|w_{x} \right|^{2}-\left|w_{y} \right|^{2})~\sim |\bm{F}_{x}|^{2} - |\bm{F}_{y}|^{2} = |\bm{F}_{x}|^{2}$.

One of the typical results from the phase interference is the harmonic structure in the scattered photons' spectrum as shown in Fig.~\ref{Fig_spec}:
with the full QED calculations, clear harmonic structure appears in the spectrum $\ud P_{x}/\ud s$ of the scattered photons in the same polarization as the laser pulse around the harmonic edges $s_{n,e}=2n\eta/(2n\eta+1+\xi^{2}/2)$ (see the black dashed lines in Fig.~\ref{Fig_spec} for $n=1,~2,~3$).
These harmonic peaks can be regarded as the result of the constructive interference between contributions from different phase points in the background field, see the phase distribution along the black dashed lines in Fig.~\ref{Fig_Ds_phi} (a), and see also the comparison with the LCFA result in Fig.~\ref{Fig_spec} in which only contributions from local fields are included~\cite{BenPRA2020}.
We point out that for a laser with longer duration, the harmonic peaks would become sharper due to the regular interference between a sequence of single-cycle pulses~\cite{LMA063110},
and for a laser with shorter duration, the harmonic peaks would be smoothed due to the ponderomotive broadening from the pulse envelope~\cite{PRA033402,PRD036018}.
However, for the photons' spectrum in the perpendicular polarization, there is no obvious harmonic structure, and the full QED and LCFA results give the similar monotone spectra, as shown in~Fig.~\ref{Fig_spec}, with only slight difference in their values in the relatively low-energy regime $s<0.1$, which could be attributed to the infrared divergence of the LCFA spectrum~\cite{king19a,PiazzaPRA2019}.
This monotone spectrum can also be inferred from the deconstructive interference contributions along the black dashed lines in Fig.~\ref{Fig_Ds_phi}~(b).

\section{Interference between two pulses}\label{Sec_3}
The phase interference is a nonlocal effect and could happen between all the points in the background field.
To analyze the polarization-dependence of the interference and quantify the phase interference effect straightforwardly, we consider a laser as
$a^{\mu}(\phi)=a^{\mu}_{1}(\phi)+a^{\mu}_{2}(\phi)$ comprised of \emph{two} pulses separated with the phase gap $\Delta=\phi_{2i}-\phi_{1f}$, where $\phi_{1f}$ is the ending of the first pulse $a^{\mu}_{1}(\phi)$ and $\phi_{2i}$ is the starting of the second pulse $a^{\mu}_{2}(\phi)$.
The phase integrals in~(\ref{Eq_cal}) can then be split into two parts and written formally, e.g., as~\cite{ILDERTON2020135410}
\begin{align}
\bm{F} = \bm{F}_{1} + e^{i\Phi_{f}}\bm{F}_{2}
\end{align}
where $\bm{F}_{1}$ ($\bm{F}_{2}$) denotes the field contribution from the first (second) pulse, and $\Phi_{f}=\Phi(\phi_{1f})+\Delta(1+r^2)s/[2 \eta (1-s)]$ is the accumulated phase before the starting of the second pulse with the term $\Phi(\phi_{1f})$ from the first pulse and the second term from the phase gap.
The interference between two pulses can then be simply seen as the cross terms $\bm{F}^{*}_{1}\bm{F}_{2}e^{i\Phi_{f}} + \bm{F}_{1}\bm{F}^{*}_{2}e^{-i\Phi_{f}}$ in the modulus square of $|\bm{F}|^{2}$~\cite{ILDERTON2020135410}.
Analogous expressions can be given for $I$ and $S$, and thus also for the scattering probability~(\ref{AllPhoton}).

\subsection{polarization-dependence of the interference}
\begin{figure}[t!!!]
\includegraphics[width=0.49\textwidth]{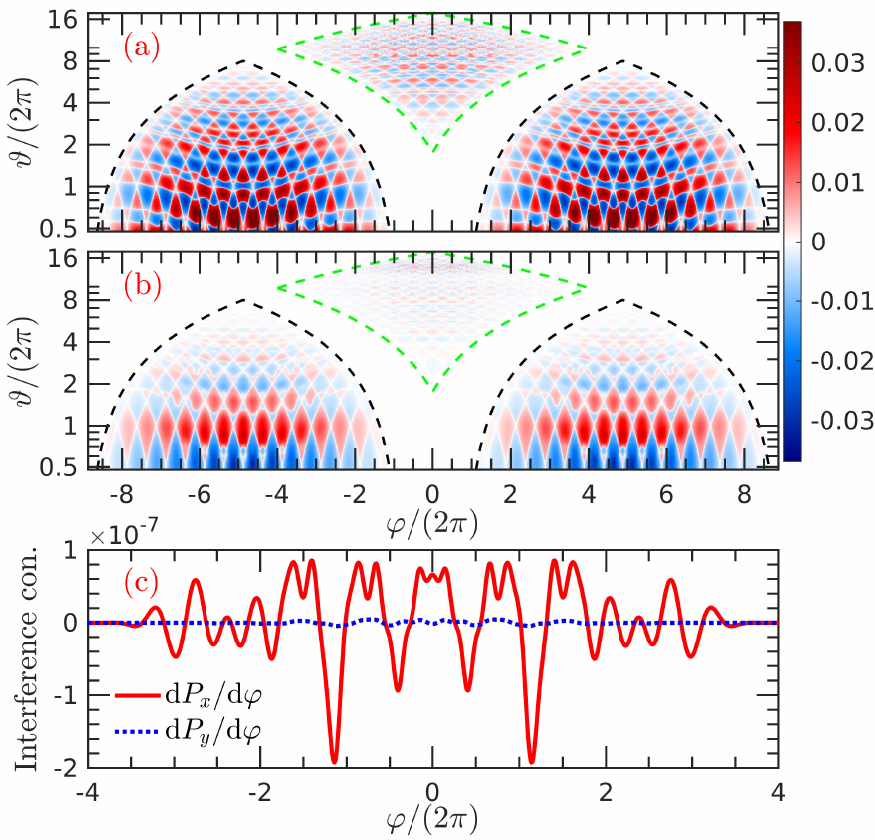}
\caption{Interference between two phase-separated pulses in the parallel polarization.
(a) $(\ud^{2}P_{x}/\ud\vphi\ud\vtheta)^{1/3}$ and (b) $(\ud^{2}P_{y}/\ud\vphi\ud\vtheta)^{1/3}$ are the phase distributions of the photons in the energy region of $0.05<s<0.25$ polarized in, respectively, the $x$ and $y$-directions.
The region surrounded by the green dashed lines is the pure contribution induced by the interference between the two pulses, and the contributions from each single pulse are surrounded by the black dashed line and the $x$-axis.
(c) plots the pure interference contribution $\ud P_{x,y}/\ud\vphi$ in (a) and (b) varying with the average phase.
The two pulse have the identical profile and are polarized along the $x$-direction with the phase gap $\Delta=3.5\pi$, and the other parameters are the same as those in Fig.~\ref{Fig_Ds_phi}.
}
\label{Fig_Aver_Intf_para}
\end{figure}

We first consider two identical pulses linearly polarized along the same direction as $a^{\mu}_{1}(\phi-\Delta/2 - N\pi)=a^{\mu}_{2}(\phi+\Delta/2 + N\pi)=m\xi\epsilon^{\mu}_{x}f(\phi)$, where $\Delta=3.5\pi$, and the first (second) pulse ends (starts) at $\phi_{1f}=-\Delta/2$ ($\phi_{2i}=\Delta/2$).
Figs.~\ref{Fig_Aver_Intf_para} (a) and (b) show the phase distributions $\ud^{2}P_{x,y}/\ud\vphi\ud\vtheta$ of the photons in the energy region of $0.05<s<0.25$ and polarized parallel and perpendicularly to the laser field, respectively.
As shown, the phase distributions can be split into three regions: the single-pulse contribution from the first pulse at $\vphi<\phi_{1f}$ and that from the second pulse at $\vphi>\phi_{2i}$, surrounded by the black dashed lines and $x$-axis,
and the pure interference contribution between the two pulses surrounded by the green dashed lines.
We can see clearly the dependence of the interference effects on the relative polarization of the laser pulse and scattered photon:
For the single-pulse contribution, the effective region of interference phase $\vtheta$ for the scattered photon in the parallel polarization as the laser field in Fig.~\ref{Fig_Aver_Intf_para} (a) is much broader than that for the perpendicularly polarized photons Fig.~\ref{Fig_Aver_Intf_para} (b), similar as the presentation in Fig~\ref{Fig_Ds_phi},
and the pure interference part in Fig.~\ref{Fig_Aver_Intf_para} (a) for the parallel polarized photon is much stronger than that in Fig.~\ref{Fig_Aver_Intf_para} (b) for the photons in the perpendicular polarization, see also the comparison in Fig.~\ref{Fig_Aver_Intf_para} (c) for the interference contribution $\ud P_{x,y}/\ud\vphi$ in each polarization.

\begin{figure}[t!!!]
\includegraphics[width=0.49\textwidth]{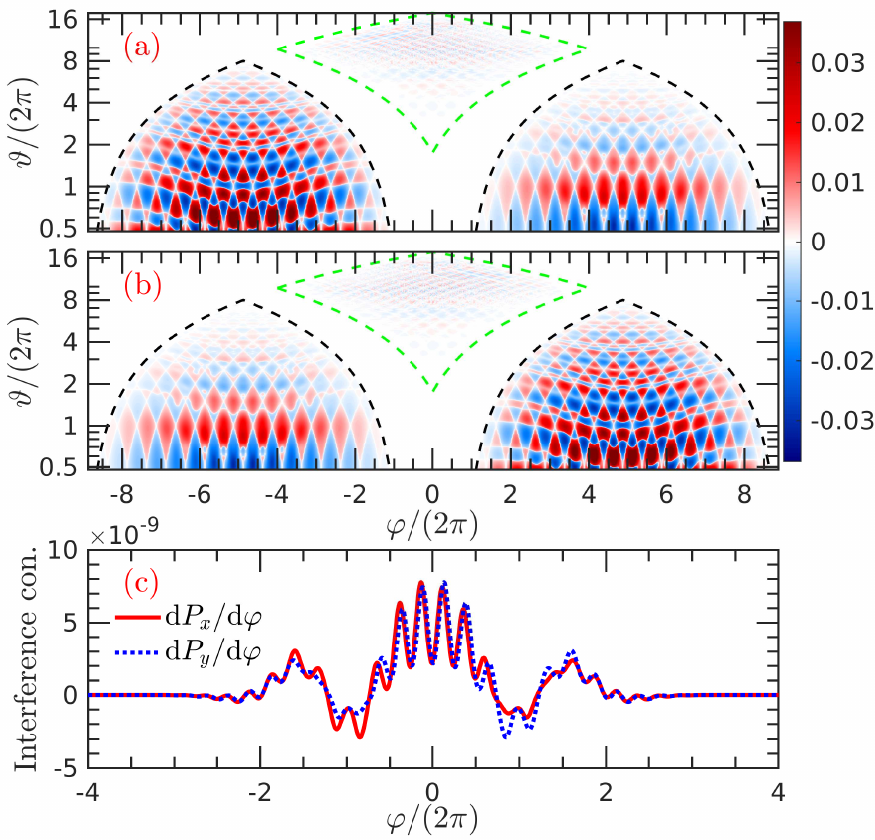}
\caption{Interference between two phase-separated pulses in the perpendicular polarization.
(a) $(\ud^{2}P_{x}/\ud\vphi\ud\vtheta)^{1/3}$ and (b) $(\ud^{2}P_{y}/\ud\vphi\ud\vtheta)^{1/3}$ are the phase distributions of the photons in the energy region of $0.05<s<0.25$ polarized in, respectively, the $x$ and $y$-directions.
The region surrounded by the green dashed lines is the interference contribution induced by the two pulses, and the contributions from each single pulse are surrounded by the black dashed line and the $x$-axis.
(c) presents the pure interference contribution $\ud P_{x,y}/\ud\vphi$ in (a) and (b) varying with the average phase.
The first (second) pulse is polarized along the $x$ ($y$) direction, and the other parameters are the same as those in Fig.~\ref{Fig_Aver_Intf_para}.}
\label{Fig_Aver_Intf_perp}
\end{figure}

In Figs.~\ref{Fig_Aver_Intf_perp} (a) and (b), we present the interference between two pulses polarized perpendicularly to each other as
$a^{\mu}_{1}(\phi)=m\xi\epsilon^{\mu}_{x}f(\phi+\Delta/2 + N\pi)$ and $a^{\mu}_{2}(\phi)=m\xi \epsilon^{\mu}_{y}f(\phi-\Delta/2 - N\pi)$, where $\epsilon^{\mu}_{y}=(0,0,1,0)$.
Similar as cases with the parallel polarization, the contributions to the scattering process can also be split into three phase regions, while with the relative importance of the different region changing with the pulse's polarization.
The first pulse polarized along the $x$-direction dominates the scattering of the $x$-polarized photons in Fig.~\ref{Fig_Aver_Intf_perp} (a) and the $y$-polarized photons in \mbox{Fig.~\ref{Fig_Aver_Intf_perp} (b)} come dominantly from the second pulse as it is polarized in the same direction.
The pure interference parts in Fig.~\ref{Fig_Aver_Intf_perp} (a) and (b) are almost the same, see the comparison in Fig.~\ref{Fig_Aver_Intf_perp} (c).
The strength of the interference between two perpendicularly polarized pulses is much weaker than that in Fig.~\ref{Fig_Aver_Intf_para} (a) for the photons polarized parallel to the pulses' polarization, due to $\bm{F}^{*}_{1}\bm{F}_{2}= \bm{F}_{1}\bm{F}^{*}_{2}=0$ for two perpendicularly polarized pulses, and in the similar level as that in \mbox{Fig.~\ref{Fig_Aver_Intf_para} (b)} for the photons in the perpendicular polarization. 

\subsection{Interference-induced observables}
We point out that for the total scattering probability, the interference contribution induced by the two pulses is much smaller than the independent contribution from each single pulse, while this interference could significantly modulate the spectral properties of the outgoing photons, which could be used as the potential experiment observables to signalize this interference effect.

In Fig.~\ref{Fig_spec}, we also present the energy spectra of the scattered photons in the double-pulse laser with the parallel polarization along the $x$ direction, and for comparison are the double-pulse spectra divided by a factor of $2$.
As shown, the interference between two phase-separated pulses reshapes the energy spectrum $\ud P_{x}/\ud s$ of the scattered photons in the parallel polarization as the laser, while the spectral shape of the photons $\ud P_{y}/\ud s$ in the perpendicular polarization is almost unchanged.
It is noteworthy that the spectral reshaping in $\ud P_{x}/\ud s$ appears dominantly around the first harmonic edge $s_{1,e}$, leading to the slight split of the first harmonic peak. 

\begin{figure}[t!!!]		
\includegraphics[width=0.49\textwidth]{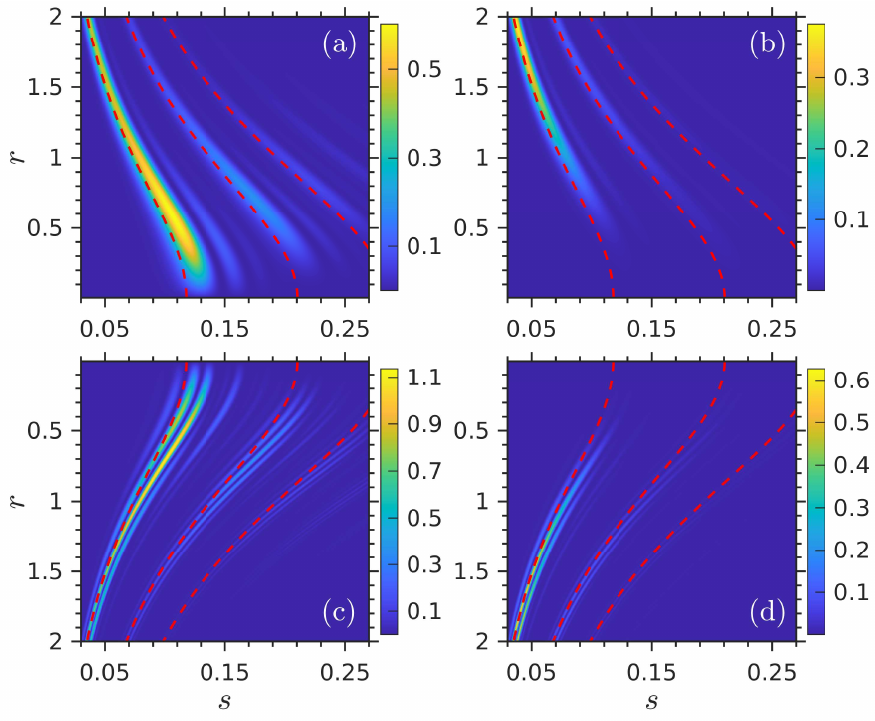}
\caption{Energy-momentum distribution $\ud^{2}P_{x,y}/\ud s~\ud r$ of the photons linearly polarized in the $x-$ (left panels) and $y-$ (right panels) directions, scattered in the laser field with one pulse (upper panels) and two pulses (bottom panels).
The two pulses are identical, linearly polarized in the $x$ direction, and separated with the phase gap $\Delta=3.5\pi$.
For comparison, the two pulse results are given by dividing a factor of $2$.
Red dashed lines are the harmonic lines $s_{n}(r) = 2n\eta/(2n\eta+1+\xi^{2}/2+r^{2})$ for $n=1,~2,~3$.
The same other parameters as in Fig.~\ref{Fig_spec} are used.}
\label{Fig2_double}
\end{figure}

This spectral reshaping can be attributed to the modulation of the energy-momentum distribution $\ud^{2}P_{x}/\ud s \ud r$ of the scattered photons in Fig.~\ref{Fig2_double}. 
For the single-pulse results in Figs.~\ref{Fig2_double} (a) and (b), the photons are scattered with the energy $s$ and transverse momentum $r$ around the harmonic line $s_{n}(r) = 2n\eta/(2n\eta+1+\xi^{2}/2+r^{2})$~\cite{MRE0196125}, see the red dashed lines for $n=1,~2,~3$ from left to right.
For the same order of harmonic, the photons polarized parallel to the laser field in Fig.~\ref{Fig2_double} (a) are scattered dominantly within a much narrower region of transverse momentum $r$ than that for the photons polarized perpendicular to the laser field in \mbox{Fig.~\ref{Fig2_double} (b)}. 
After integrating over the transverse momentum, the harmonic peaks are thus risen up in the energy spectrum of the parallel polarized photons around the harmonic edge $s_{n,e}$, but smoothed out in the spectrum of the perpendicularly polarized photons as shown in Fig.~\ref{Fig_spec}.
For the double-pulse results in Figs.~\ref{Fig2_double} (c) and (d), clear interference fringes appear in $\ud^{2}P_{x,y}/\ud s \ud r$ along the harmonic lines $s_{n}(r)$, and crucially, the separation between the fringes decreases gradually with the increase of the transverse momentum $r$.
After integrating over transverse momenta, the interference fringes at larger $r$, contributing to smaller energy $s$, would be smoothed out and only those at smaller $r\to0$ can sustain in the photon spectrum, thus leading to the spectral reshaping in $\ud P_{x}/\ud s$ around the first harmonic edge $s_{1,e}$.
Because of the larger transverse momenta in Fig.~\ref{Fig2_double} (d) for the perpendicularly polarized photons, there's no clear interference signal sustained in their final energy spectrum.
We also note sub-fringes between the harmonic lines in Fig.~\ref{Fig2_double} (a) and (b), which is also interference result induced by the finite pulse envelope~\cite{PRD036018,Tang:2021qht}.


\begin{figure}[t!!!]
\includegraphics[width=0.48\textwidth]{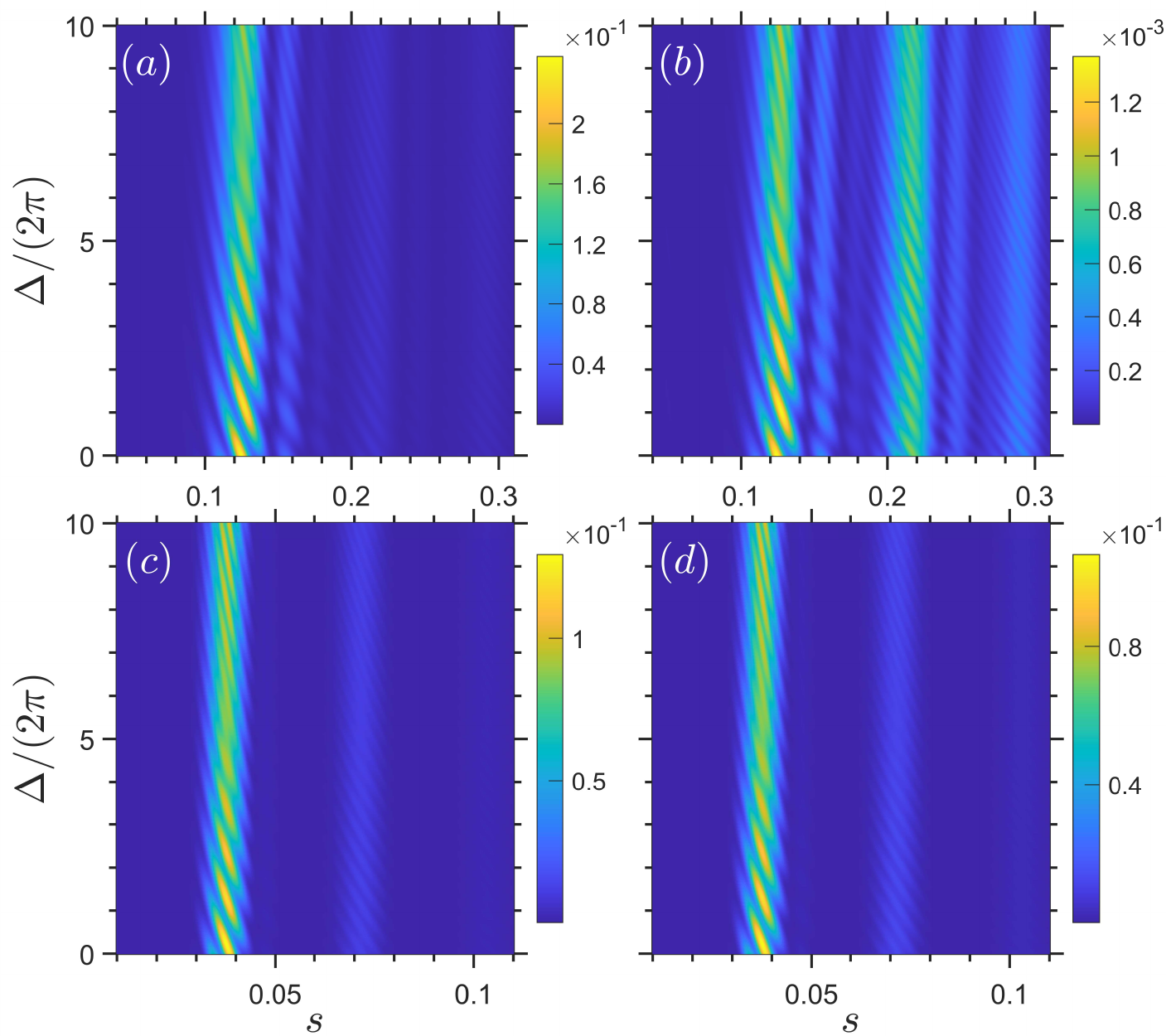}
\caption{Variation of the energy spectra $\ud P_{x,y}/\ud s$ with the increase of the phase gap $\Delta$ between the two pulses in the parallel polarization, for the photons linearly polarized in the $x$ (left panels) and $y$ (right panels) directions.
In (a) and (b), the energy spectra of the photons with the transverse momentum $r<0.3$, corresponding to the scattering angle $\theta<18~\mu\trm{rad}$, are plotted, and in (c) and (d) with $1.9<r<2.0$, corresponding to the scattering angle $115~\mu\trm{rad}<\theta<121~\mu\trm{rad}$.
The same parameters as in Fig.~\ref{Fig2_double} are used.}
\label{Fig_delta}
\end{figure}

From Figs.~\ref{Fig2_double}~(c) and~(d), we can infer that to manifest the interference signal in the photon energy spectra, one may integrate partially the double differential spectra over finite transverse momentum, i.e., accepting the scattered photons in the specified direction.
Fig.~\ref{Fig_delta} shows the energy spectra $\ud P_{x,y}/\ud s$ of the scattered photons within a narrow region of transverse momentum with the variation of the phase gap $\Delta$.
As shown in Figs.~\ref{Fig_delta}~(a) and (b) for the photons with the transverse momentum $r<0.3$, corresponding to the scattering angle $\theta<18~\mu\trm{rad}$, clear interference fluctuations appear in both the energy spectra of photons polarized parallel [Fig.~\ref{Fig_delta}~(a)] and perpendicular [Fig.~\ref{Fig_delta}~(b)] to the laser field, even though the scattering probability for perpendicularly polarized photons in Fig.~\ref{Fig_delta} (b) is much smaller than that in Fig.~\ref{Fig_delta} (b) for parallel polarized photons.
For photons with larger scattering angles, a narrower region of transverse momentum should be integrated over for the manifestation of interference signal in their energy spectra as shown in \mbox{Figs.~\ref{Fig_delta}~(c) and (d)} for $1.9<r<2.0$ and $115~\mu\trm{rad}<\theta<121~\mu\trm{rad}$.
We also note in Fig.~\ref{Fig_delta} that the sharp spectral fluctuations at $\Delta\to 0$ becomes smoother with the increase of phase gap, and becomes almost indistinguishable at $\Delta> 20\pi$. 
To manifest the spectral interference signal at larger phase gap $\Delta$, a narrower region of transverse momentum should be integrated over~\cite{ILDERTON2020135410}.  %

To understand how the double-pulse interference effect modulate the photon spectra, we can write the two-pulse interference analytically.
For the pulse profile in Fig.~\ref{Fig2_double}, the accumulated phase can be given explicitly as
\begin{align}
\Phi_{f}=\frac{s}{2 \eta (1-s)}[(1+r^2)(2N\pi+\Delta) + 3N\pi\xi^{2}/8 ]\,.
\label{Eq_Inter_phase}
\end{align}
We point out that for few-cycle pulses, i.e., $\int \bm{a}_{1}\ud \phi \neq0$, the accumulated phase would depend also on the azimuthal angle of the scattered photons.
For the two identical pulses, \emph{i.e.},~$\bm{F}_{1}=\bm{F}_{2}$, the energy-momentum distribution in the double-pulse case can be related to that in the single-pulse case as
\begin{align}
\left. \frac{\ud^{2} P_{x,y}}{\ud s~\ud r}\right|_{\trm{double}}= 2(1+\cos\Phi_{f})\left.\frac{\ud^{2}P_{x,y}}{\ud s~\ud r}\right|_{\trm{single}}\,,
\label{Eq_interf}
\end{align}
in which the interference between the two pulses can be clearly seen as the factor $\cos\Phi_{f}$.
By dropping the interference factor $\cos\Phi_{f}$, i.e., neglecting the interference effect, the emission spectrum would go back to the incoherent sum over the contribution from each pulse.

The interference fringes shown in Figs.~\ref{Fig2_double}~(c) and~(d) are actually resulting from the variation of the accumulated phase $\Phi_{f}$: $\Phi_{f}=(2j+1)\pi$ and $\Phi_{f}=2j\pi$ correspond respectively to the minima and maxima in the energy-momentum distributions, where $j$ is an arbitrary integer. 
With the increase of photon energy $s$ and transverse momentum $r$,
the variation of $\Phi_{f}$, i.e., $\ud \Phi_{f}/\ud s$ and $\ud \Phi_{f}/\ud r$, would become more rapid, resulting in the tighter interference fringes in the regimes of larger $s$ and $r$ in Figs.~\ref{Fig2_double}~(c) and (d).
As we can also see in~(\ref{Eq_Inter_phase}), the variation of $\Phi_{f}$ would be further pronounced with the increase of the phase gap $\Delta$ between the two pulses.
This leads to the more frequent fluctuations in the energy spectrum at larger phase gap $\Delta$ shown in Fig.~\ref{Fig_delta}.

We also point out that as the interference factor in~(\ref{Eq_interf}) is independent on the photon polarization, the similar interference fringes would appear in the spectra of the total scattered photons as those in Figs.~\ref{Fig2_double} (c) and (d), and the interference between two laser pulses with parallel polarization would not change the energy-momentum distribution of the photon polarization degree.
\ST{However, the different distribution of the polarized photons' transverse momentum in~Fig.~\ref{Fig2_double} leads to the different interference manifestation in the polarized energy spectra in Fig.~\ref{Fig_spec}, and could thus result into slight difference in the photons' polarization spectrum.
}

As the interference between two perpendicularly polarized pulses is much weaker than that between two parallel polarized pulses,
there is only weak fringes appearing in the total energy-momentum distribution of the scattered photons, 
and thus no spectral reshaping in their energy spectra. 
If the two pulses are identical and polarized perpendicularly in the $x$ and $y$ direction respectively, they contribute to the scattering process in the same level and symmetrically, e.g., as $F_{1x}(r_{x},r_{y}) = F_{2y}(r_{y},r_{x})$.
Therefore, the energy-momentum distributions $\ud^{2} P_{x,y}/\ud s\ud r$ of the photons linearly polarized in the $x$ and $y$ direction are equal after the azimuthal integral as \mbox{$\int \ud\psi (\left|w_{x} \right|^{2}-\left|w_{y} \right|^{2})~\sim \int \ud\psi (|\bm{F}_{x}|^{2} - |\bm{F}_{y}|^{2}) = 0$}, giving thus $\ud P_{x}/\ud s = \ud P_{y}/\ud s$.


\section{Conclusion}~\label{Sec_4}
We investigate the phase interference effects in the nonlinear Compton scattering in the collision between an high-energy electron and a laser field with linear polarization,
and reveal that the importance of the interference effects on the scattered photons depends sensitively on the relative polarization of the scattered photons and laser pulse. 
For scattered photons in the parallel polarization to the laser field, the effective distance between two interfered phase points is much larger than that for perpendicularly polarized photons.

The spectral properties of the scattered photons could be significantly modulated by the phase interference effect.
For a single-pulse laser with long-enough duration, the constructive interference between different field points results in the harmonic peak in the energy spectrum of the parallel polarized photons.
For the laser with two phase-separated pulses, the interference between the two pulses could modulate the energy-momentum distribution of the scattered photons, and reshape the energy spectrum of the scattered photons in the parallel polarization, while the spectral shape of the photons in the perpendicular polarization is almost same as that in the single-pulse case.

The interference-induced modulation into the photons' energy-momentum distribution depends sensitively on the scattering angle of the photons.
By narrowing the angular acceptance of the scattered photons, the interference fluctuation in the photons' energy spectrum can be effectively amplified and sustains for an experiment-realizable phase separation between two pulses.
\ST{To observe the harmonic spectral reshaping, the laser pulses with multiple cycles should be employed in the intermediate intensity regime of $\xi\sim O(1)$, in which the upcoming experiments such as E320 and LUXE are designed.}

\ST{To observe robust interference signals in experiments, one should consider not a single electron, but a bunch of energetic electrons, which could be modeled by incoherently integrating the differential probability with the momentum distribution of the electron bunch~\cite{PRA2016Angioi}.
Therefore, the energy spread of the electron bunch must be narrow enough in order to resolve the interference-induced spectral fluctuations.
For the electron bunch with high density and broad momentum distribution, the coherent emission from the bunch of electrons should also be incorporated in future analysis~\cite{PRL010402,PRD035048}.
}

\section{Acknowledgments}
The authors acknowledge the support from the National Natural Science Foundation of China, Grant No.~12575261.
The work was carried out at Marine Big Data Center of Institute for Advanced Ocean Study of Ocean University of China.

\bibliographystyle{apsrev}

\end{document}